\documentclass[pra,aps,twocolumn,amsmath,amssymb]{revtex4}
\usepackage{graphicx}
\pdfoutput=1
\usepackage{amsmath}
\usepackage{color}
\usepackage{float}

\begin{document}

\title{Observation of hysteretic phase switching in silicon by piezoresponse force microscopy}
\author{Jagmeet S. Sekhon}
\author{Leena Aggarwal }
\author{Goutam Sheet}
\email{goutam@iisermohali.ac.in}
\affiliation{Department of Physical Sciences,  
Indian Institute of Science Education and Research, Mohali
Punjab, India, PIN: 140306}

\begin{abstract}
We report the observation of $180^o$ phase switching on silicon wafers by piezo-response force microscopy (PFM). The switching is hysteretic and shows remarkable similarities with polarization switching in ferroelectrics. This is always accompanied by a hysteretic amplitude vs. voltage curve which resembles the ``butterfly loops" for piezoelectric materials. From a detailed analysis of the data obtained under different environmental and experimental conditions, we show that the hysteresis effects in phase and amplitude do not originate from ferro-electricity or piezoelectricity. This further indicates that mere observation of hysteresis effects in PFM does not confirm the existence of ferroelectric and/or piezoelectric ordering in materials. We also show that when samples are mounted on silicon for PFM measurements, the switching properties of silicon may appear on the sample even if the sample thickness is large.

\end{abstract}

\maketitle

Owing to its semiconducting behaviour, resilience against high temperature and high electrical power, silicon has become one of the most popular substrates for wide variety of systems ranging from thin solid films to bio-materials.\cite{1,2,3} Such systems grown/mounted on silicon are studied by diverse characterization and measurement probes. One of the important measurement techniques where silicon is ubiquitously used as a substrate in piezoresponse force microscopy (PFM).\cite{2,3,4} In PFM spectroscopy, a conducting AFM (atomic force microscope) cantilever is brought in touch with a sample and the electro-mechanical interaction between the cantilever and the sample is studied to determine how the sample responds to a sweeping dc electric field applied on the cantilever. An ac signal rides on the sweeping dc field for tracking the sample characteristics as the dc field sweeps.\cite{4,5} In PFM spectroscopy $180^o$ phase switching with voltage and a hysteresis in the phase vs. dc voltage plot is traditionally considered to be a signature of polarization switching thereby confirming the material under study to be ferroelectric. Furthermore, a hysteretic amplitude vs. dc voltage curve (traditionaly known as a ``butterfly loop") is considered to be the hallmark of piezoelectricity.\cite{5} Silicon has been used in the past as a very popular substrate for searching potential ferroelectrics/piezoelectrics by PFM in solid semiconducting films\cite{6}, ferroelectric films,\cite{7} soft materials\cite{3}, bio-materials\cite{8,9} etc. Silicon is widely used as a substrate for PFM because it is thought that silicon is not ferroelectric and therefore it does not contribute to phase-switching. 

In the past the effect of the application of a large electric field by an AFM tip on the surface of silicon in air was studied. In such studies, it was observed that the silicon surface undergoes an electrochemical reaction resulting in nano-scale structures of silicon oxide on the surface under the AFM tip.\cite{10,11,12,13,14,15} However, the electrical response of silicon was not studied by PFM spectroscopy and the role of the substrate in PFM measurements on materials mounted on silicon remained unknown. 

In this Letter, we report PFM measurements on silicon wafers in air. The silicon wafers exhibit clear and strong hysteresis as well as ``butterfly loops". Nano-structures of silicon oxide grow under the PFM tip during the hysteresis measurements due to the electrochemical reaction initiated by the applied measurement potential. From the measurements on thick non-ferroelectric samples mounted on silicon, we show that the observed hysteresis on silicon in air may contribute significantly to the response from the sample mounted on it. This might give rise to misleading results about the ferroelectric/piezoelectric properties of samples mounted on silicon.

For the PFM spectroscopic measurements, a conducting cantilever was brought in contact with the surface of a $5 mm$ x  $5 mm$ piece of silicon. An AC excitation voltage ($V_{ac}$) of $10V$ was applied to the cantilever. The amplitude response of the cantilever as a function of frequency was recorded in order to characterize the in-contact resonance of the cantilever. As shown in Figure 1(a), the in-contact resonance frequency varied between 280 to 300 kHz. The measurements were carried out at the resonance frequency in order to achieve higher sensitivity and for simultaneously obtaining spectroscopic information regarding strain and dissipation. The entire work was done in dual AC resonance tracking (DART) mode of PFM.\cite{16} $V_{ac}$  was kept constant at $10V$ during all the measurements. 

In Figure 1(b), we show the hysteresis in phase ($\phi$) vs. dc bias ($V_{dc}$) plots for different ranges of $V_{dc}$. At a relative humidity of $27\%$, we clearly observe hysteresis loops for all ranges of $V_{dc}$, where the phase switches by nearly $180^o$ with a coercive voltage of $18-20 V$. A phase switching of $180^o$ is generally believed to originate from polarization switching in ferro-electrics. The coercive voltage in this case depends on the maximum $V_{dc}$ applied. Coercive voltage increases by approximately $10\%$ as the maximum applied $V_{dc}$ increases from $40 V$ to $60 V$. The overall shape of the hysteresis loop significantly depends on the ambient relative humidity. At higher humidity levels (typically for $R_H > 40\%$), we also observe a dip near the switching bias in the hysteresis loop. Figure 1(c) shows the hysteresis data obtained at $R_H = 48\%$, where the dip is very large. The amplitude ($A$) vs. $V_{dc}$ curve (shown in Figure 1(d)) is also hysteretic and the shape of the loop strongly resembles the so-called ``butterfly loop" normally observed in piezoelectrics. The amplitude ($A$) is directly related to the local strain experienced by the cantilever. In case of piezo-electric materials the hysteresis in strain vs. voltage originate from the dynamics of the piezoelectric domains in the material under an applied electric field\cite{17}. 
\begin{figure}
\begin{center}
\includegraphics[width=0.5\textwidth]{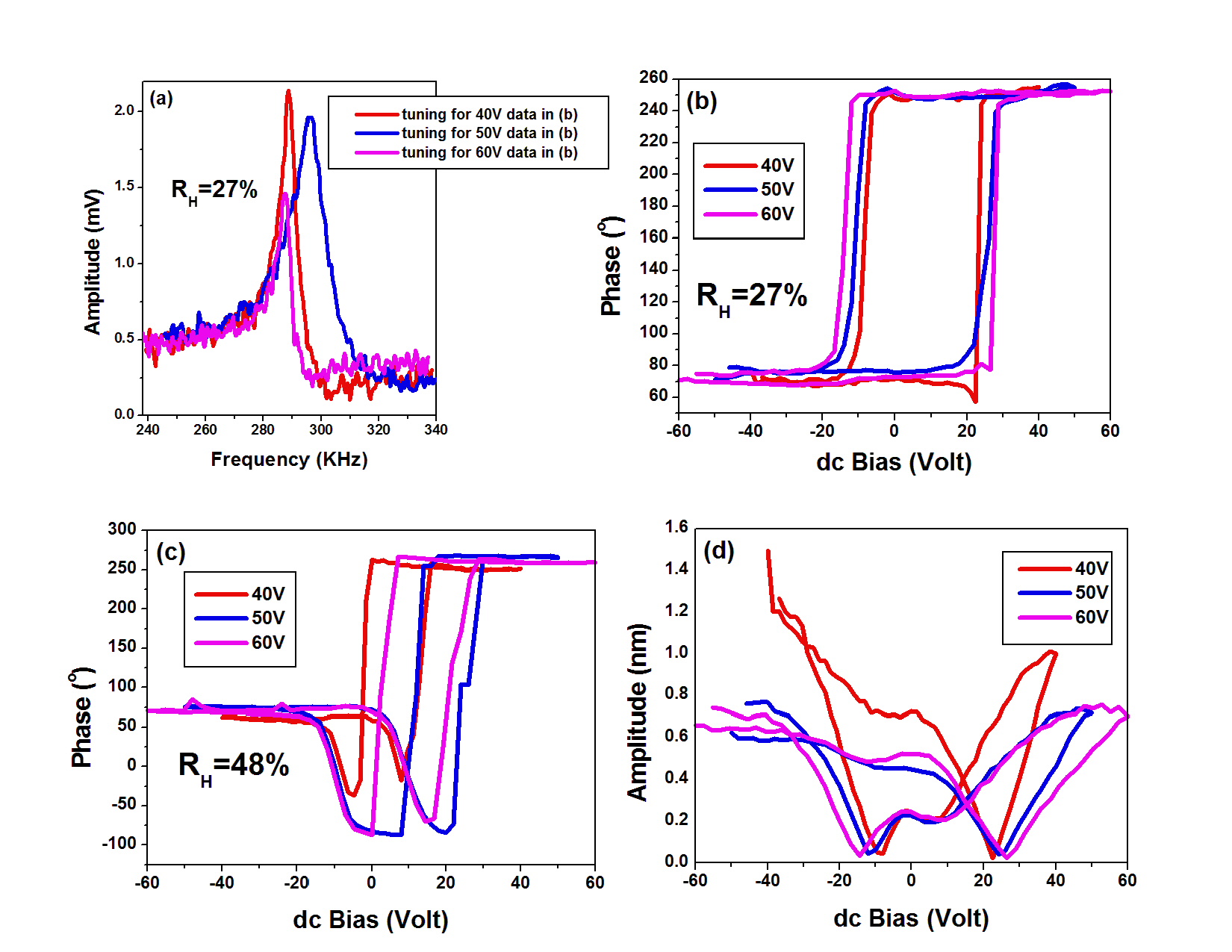}
\end{center}
\caption{``Piezo-response"-like signal observed on silicon: (a) Tuning of the conducting AFM cantilever in the PFM mode prior to spectroscopic measurements, (b) Phase vs. dc-Bias data (at relative humidity $R_H = 27\%$) showing hysteresis for the maximum dc- voltage of $40V$ (red), $50V$ (blue), and $60V$ (magenta), (c) Phase vs. dc-Bias data (at relative humidity $R_H = 48\%$) showing hysteresis for maximum dc- voltages of $40V$ (red), $50V$ (blue), and $60V$ (magenta), (d) Amplitude vs. dc-Bias data showing ``hysteresis loops" for maximum dc- voltages of $40V$ (red), $50V$ (blue) and $60V$ (magenta). 
}
\end{figure}

After every spectroscopic measurement we imaged the topography in non-contact mode and observed that nanometer size structures were grown on the points where the measurements were carried out. The overall size of the nano-structures varied as the maximum range of $V_{dc}$ varied. In Figure 2(a), we show the topographic image of an area where a number of measurements were done with different ranges of $V_{dc}$. The nano-structures grown during the spectroscopic measurements for different $V_{dc}$ are clearly visible in the figure.\cite{18} In Figure 2(b) we show the height and width of the nano-structures grown corresponding to the line-cuts drawn in Figure 2(a). In the insets of figure 2(b) and 2(d) we also show how the heights vary with the maximum $V_{dc}$ range applied. 
It is seen that the height of the nano-structures grown during spectroscopy are $90nm$, $190nm$, $210nm$, $470nm$ for the maximum $V_{dc}$ ranges of $80V$, $90V$, $100V$, and $110V$ respectively.

\begin{figure}
\begin{center}
\includegraphics[width=0.5\textwidth]{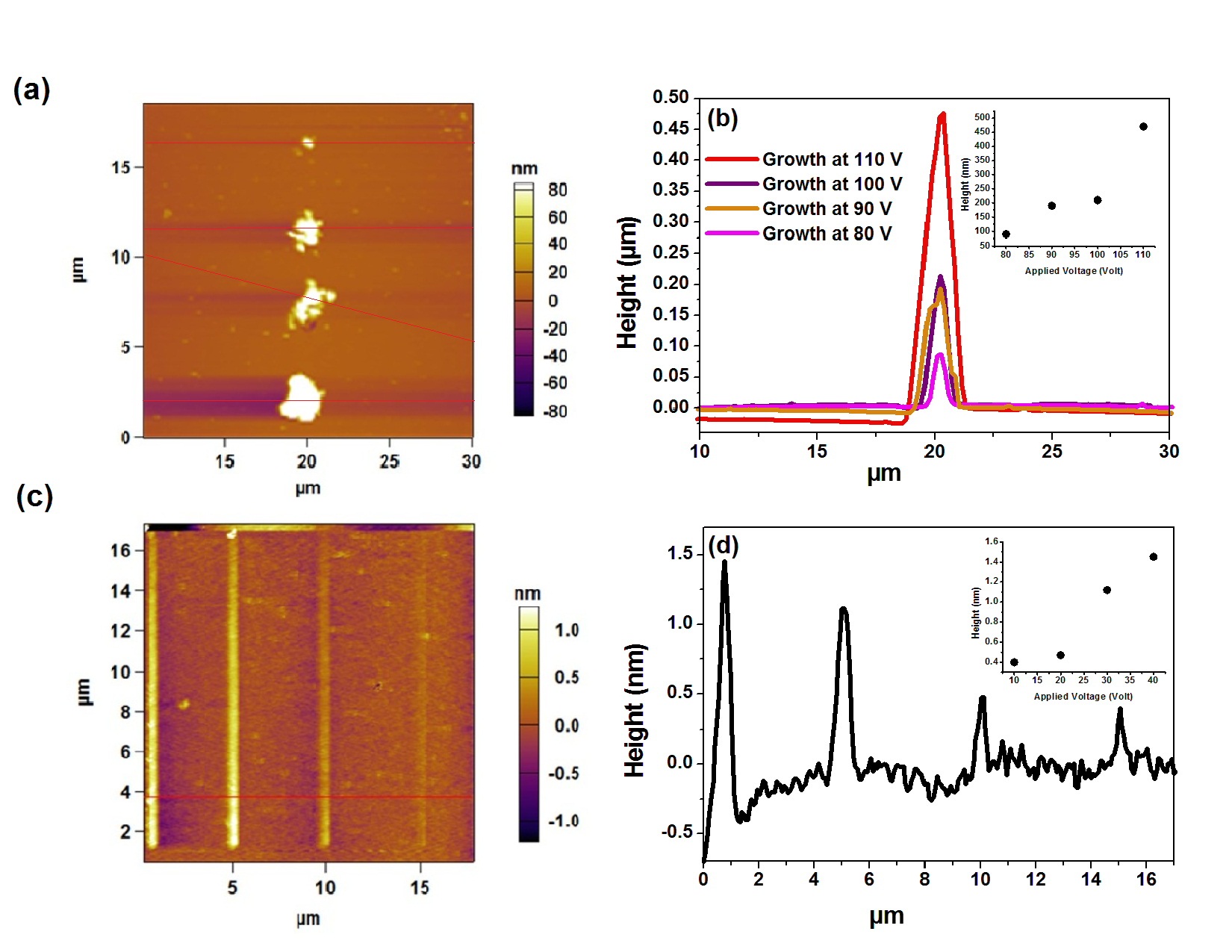}
\end{center}
\caption{(a) Topographic image of nano-structures grown on silicon during spectroscopic measurements with maximum bias of $80V$, $90V$, $100V$, $110V$, (b) height variation of the nano-structures grown during measurements at $80V$ (magenta), $90V$ (orange), $100V$ (purple), $110V$ (red)  corresponding to the line-cuts shown in (a),(c) Topographic image of  nanowires lithographically written on silicon by the AFM tip at $10V$, $20V$, $30V$, and $40V$, (d) height variations of nanowires corresponding to the line-cut in (c).}
\end{figure}

In order to understand the possible role of applied bias on the growth of nano-structures during spectroscopic measurements, we have attempted to grow nano-structures on the silicon surface by usual lithographic technique. In this technique simply a dc-voltage is applied to the tip and the tip is scanned along a designed geometrical shape on the sample surface.\cite{19} If the sample material is electrochemically active, the applied voltage will write a nano-structure on the sample as the tip scans over it. In contrast, during the spectroscopic measurements, the tip remains in contact with the sample at a particular point and the voltage is scanned between $-V_{dc}$ and $+V_{dc}$. In Figure 2(c), we show four parallel nano-wires lithographically grown on the same silicon wafer by a conducting tip with applied voltages of $10V$, $20V$, $30V$, and $40V$, respectively. From the line profile corresponding to the line-cut shown in the image (Figure 2(d)), it is clear that height and the width of the lithographically written nano-wires are dependent on the voltage between the tip and Silicon. The height of the nano-wires are $0.40nm, 0.47nm, 1.12nm, 1.45nm$ at $10V, 20V, 30V,$ and $40V,$ respectively. The growth of such nano-wires on silicon surfaces was investigated\cite{10,11,13,20,21}, and was attributed to electro-chemical reactions on the silicon surface in presence of moisture. From various analytical tools it was inferred that the chemical composition of these structures were primarily silicon oxide\cite{22}. Therefore, it is rational to conclude that the nano-structures grown during the spectroscopic measurements that we have carried out are also primarily made up of silicon oxide formed due to the application of a high measurement voltage in the presence of moisture.

In the context of electrochemical strain microscopy (ESM), it has been demonstrated that the growth of nano-structures underneath a conducting cantilever might give rise to a deflection of the cantilever due to the electrochemical strain developed during the reaction.\cite{23} This might also lead to a hysteresis in the phase and amplitude as we observe in the case of silicon. In case of silicon, since the shape of the hysteresis loop depends on the relative humidity of the ambience during the measurements, it can be inferred that the electrochemical processes are indeed playing prominent role in generating the hysteresis effects. From the AFM imaging and PFM spectroscopy alone it is impossible to comment on the reaction dynamics under the tip. From the visual inspection of the hysteresis data it is clear that the reaction dynamics is different at different humidity levels. While the hysteresis itself may originate from some electrochemical reaction that leads to the nano-structure growth, the dip in the phase vs. voltage curve at high humidity might arise from a secondary chemical reaction favoured by high level of moisture. This is somewhat similar to what is observed in the current($I$) vs. voltage ($V$) response in cyclic voltametry of electrochemical processes with liquid electrolytes.\cite{24} It should be noted that even in the case of typical cyclic voltametry, the current vs. voltage curves often show hysteresis and when more than one reactions are involved, corresponding multiple features are observed in the hystertic $I-V$ curves. 

It is usually believed if the thickness of a given sample is large enough, the contribution from the substrate in the PFM results obtained on the samples (mounted on the substrate) is not significant. This idea has been quantitatively explored in the modified effective charge model developed by Morozovska et.al.\cite{25} According to this model, the electric field induced by a charged PFM tip on the surface is given by $E_3\propto \frac{Q}{(x_3+d)^2}$, where $Q$ is the effective charge on the tip which is proportional to the applied voltage $V_{dc}$, $d$ is the distance between the tip and the equipotential surface underneath the sample and $x_3$ is the distance of the measured surface from the equipotential surface under the sample, which is usually equal to the sample thickness. It is argued that when the sample is thick i.e., $d$ is large, the effective electric field experienced by the underlying substrate is small and hence, the voltage induced dynamics of the substrate should not contribute in the spectroscopic measurements on the sample. Even though this model is widely used for analysing PFM data on new materials, it should be noted that this model has been developed for ferroelectric domain nucleation and switching. When the hysteresis appears from phenomena not related to ferro-electricity, applying this model to rule out the role of the substrate for thick samples may not be appropriate. In order to investigate the influence of the silicon substrate on PFM measurements, we mounted a $1.36 mm$ thick PCB plastic board (a known non-ferroelectric insulator) on silicon and observed clear hysteresis and ``butterfly loops" (Figure 3(a), 3(c)). Furthermore, for $R_H > 40\%$ a dip near the switching bias appeared, which is similar to the characteristic dip observed on silicon (Figure 3(b)) for $R_H > 40\%$ . When the measurement is done on the piece of PCB alone (mounted on a metal disk), some hysteretic behaviour is observed that shows identical shape at all humidity levels i.e., the dip does not appear at high humidity levels. Hence, it is clear that the dip observed in PCB mounted on silicon at higher humidity levels appears due to the contribution from the silicon substrate. Therefore, it can be concluded that when PFM measurements are performed on samples mounted on silicon, fake ferroelectric-like signal may be obtained. 

\begin{figure}
\begin{center}
\includegraphics[width=0.5\textwidth]{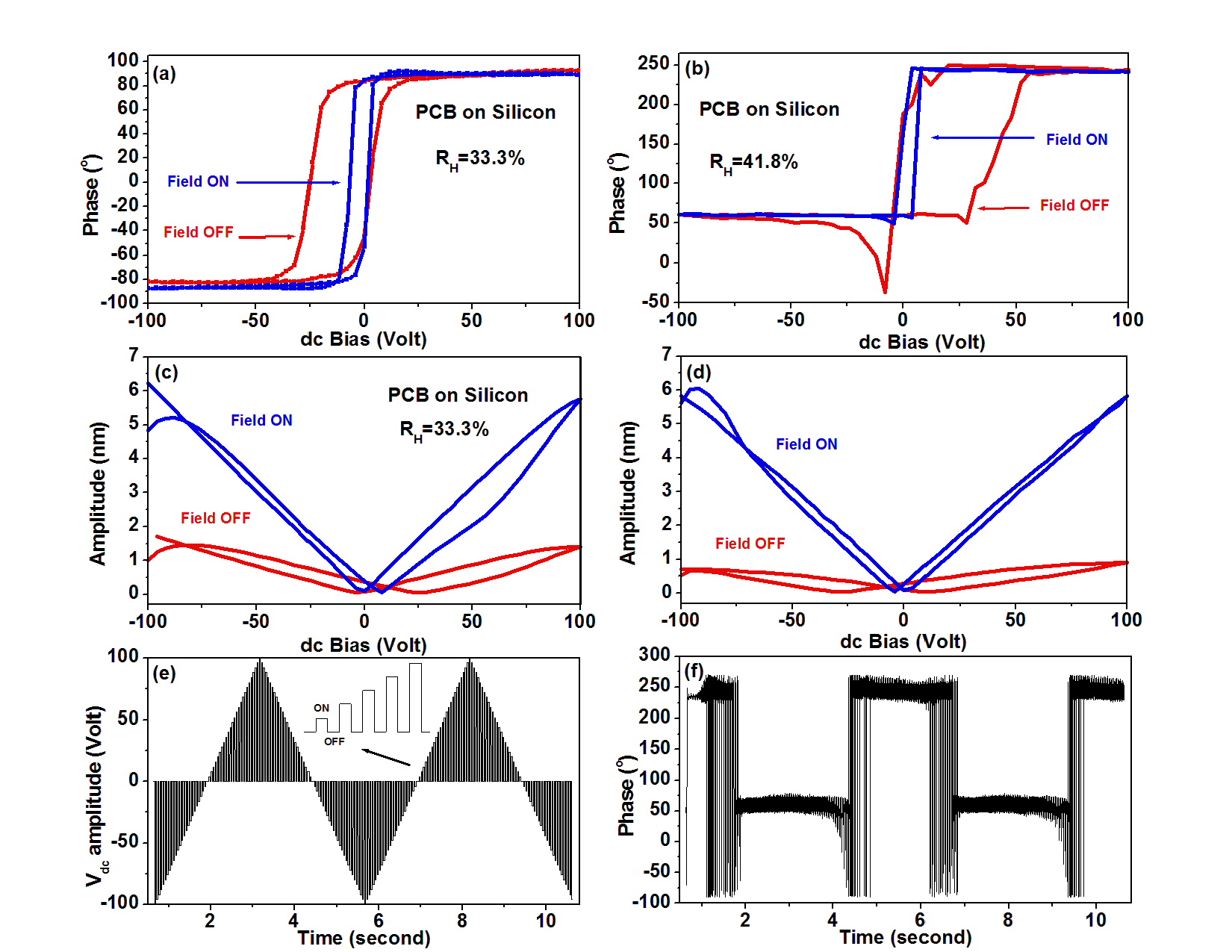}
\end{center}

\caption{ Phase vs. dc-Bias voltage in ``on" (blue) and ``off" (red) states measured on a piece of $1.36 mm$ thick PCB mounted on silicon at relative humidity of (a) $R_H=33.3\%$, (b) $R_H=41.8\%$. Amplitude vs. dc-Bias voltage measurement showing the typical ``butterfly loops" on PCB mounted on silicon at relative humidity of (c) $R_H=33.3\%$, (d) $R_H=41.8\%$. (e) Triangular pulsed signal used for biasing the tip during spectroscopic measurements, (f) typical phase-switching behaviour in time domain}
\end{figure}

It should be noted that in PFM measurements the hysteresis effects could also arise due to the capacitive coupling between the conducting tip and the sample.\cite{14} In order to mitigate this effect all the spectroscopic data reported here were measured by employing a switching spectroscopy PFM (SSPFM) protocol pioneered by Jesse et.al., where a sequence of dc voltages in triangular saw tooth form is applied between the conducting cantilever and the silicon wafer.\cite{26,27} A profile of the typical waveform applied during such measurements is shown in Figure 3(e) and the corresponding $180^0$ phase switching in time domain is shown in Figure 3(f). Bias dependent responses of phase and amplitude were recorded in the ``off" state to minimize the contribution of electrostatic interaction.\cite{3} As it is seen in Figure 3(b) and 3(d), the ``off" state $\phi$ vs. $V_{dc}$ and $A$ vs. $V_{dc}$ curves saturate at higher voltages than the curves recorded in the ``on" state. In addition, the ``coercive voltage" in the ``off" state is also seen to be higher than that in the ``on" state. This difference indicates that the role of electrostatic interaction in the observed results in the ``off" state is insignificant.

In conclusion, we have performed PFM measurements on silicon wafers and observed ferroelectric like hysteresis in phase vs. voltage and piezo-electric like ``butterfly loops" in amplitude vs. voltage curves. The overall shape of the hysteresis loops depends on the level of relative humidity in the ambience during measurements. We also observe growth of nano-structures under the cantilever during hysteresis measurements. We attribute all the observation to electrochemical processes taking place under the cantilever. When a thick non-ferroelectric sample material is mounted on silicon for PFM measurements, the hysteresis is observed on the sample which shows similar humidity dependence as silicon confirming that the observed hysteresis in such case results from silicon. This makes silicon a bad choice as substrate for PFM measurements. we also argue that observation of hysteresis loops alone in PFM is not a proof of ferroelectricity and/or piezoelectricity in materials as strikingly similar hysteresis may also arise from phenomena other than ferroelectricity and/or piezoelectricity.

We thank A. Arora and J. Bhagwathi for their help. We acknowledge fruitful discussions with R. Proksch. GS acknowledges the research grant of Ramanujan Fellowship from Department of science and Technology (DST), India for partial financial support.


\begin{thebibliography}{99}

\bibitem{1}	F. Priolo, T. Gregorkiewicz, M. Galli, T. Krauss, Nature Nanotechnology \textbf{9,} 19 (2014).
\bibitem{2} M. Lebedev, J. Akedo, Y. Akiyama, Jpn. J. Appl. Phys. \textbf{39,} 5600 (2000).
\bibitem{3} Y. Liu, Y. Zhang, M. J. Chow, N. Q. Chen, J. Li, Phys. Rev. Lett. {\bf 108}, 078103 (2012).
\bibitem{4} E. Soergel, J. Phys. D: Appl. Phys. \textbf{44,} 464003 (2011).
\bibitem{5} S. V. Kalinin, B. J. Rodriguez, S. Jesse, P. Maksymovych, K. Seal, M. Nikiforov, A. P. Baddorf, A. L. Kholkin, R. Proksch, Materialstoday \textbf{11,} 16 (2008).
\bibitem{6} C. Dubourdieu, J. Bruley, T. M. Arruda, A. Posadas, J. Jordan-Sweet, M. M. Frank, E. Cartier, D. J. Frank, S. V. Kalinin, A. A. Demkobv, V. Narayan, Nature Nanotechnology \textbf{8,} 748 (2013).
\bibitem{7} M. Narayanan M. Pan, S. Liu, S. Tong, S. Hong, B. Ma, U. Balachandran, RSC Advances \textbf{2,} 11901 (2012).
\bibitem{8} B.J. Rodriguez, S. V. Kalinin, J. Shin, S. Jesse, V. Grichko, T. Thundat, A. P. Baddorf, A. Gruverman, Journal of Structural Biology \textbf{153,} 151 (2006).
\bibitem{9} S. B. Lang, S. A. M. Tofail, A. L. Kholkin, M. Wojtas, M. Gregor, A. A. Gandhi, Y. Wang, S. Bauer, M. Krause, A. Plecenik, Scientific Reports \textbf{3,} 1 (2013).
\bibitem{10} P. M. Campbell, E. S. Snow, P. J. McMarr, Physica B \textbf{227,} 315 (1996).
\bibitem{11} J.B. Wang, M. Y. Lai, Y. L. Wang, Chinese Journal of Physics \textbf{36,} 642 (1998).
\bibitem{12} D. Wang, L. Tsau, K. L. Wang, Applied Physics Letter \textbf{65,} 1415 (1994).
\bibitem{13} M. T. Asmah, S. D. Hutagalung, O. Sidek, Journal of Physics: Conference Series \textbf{431,} 1 (2013).
\bibitem{14} J. A. Dagata, F. Perez-Murano, C. Martin, H. Kuramochi, H. Yokoyama, J. Appl. Phys. \textbf{96,} 2386 (2004)
\bibitem{15} M. Tello, R. Garcia, Appl. Phys. Lett. \textbf{79,} 424 (2001).
\bibitem{16} B. J. Rodriguez, C. Callahan, S. V. Kalinin, R. Proksch, Nanotechnology\textbf{ 18,} 475504 (2007).
\bibitem{17} S. V. Kalinin, E. Karapetian, M. Kachanov, Phys. Rev. B \textbf{70,} 184101 (2004).
\bibitem{18} A. Dehzangi, F. Larki, S. D. Hutagalung, M. G. Naseri, B. Y. Majilis, M. Navasery, N. A. Hamid, M. M. Noor, PLOSone \textbf{8,} 1 (2013).
\bibitem{19} X. N. Xie, H. J. Chung, C. H. Sow, A. T. S. Wee, Materials Science and Engineering R \textbf{54,} 1 (2006).
\bibitem{20} A. Giguere, J. Beerens, B. Terreault, Nanotechnology \textbf{17,} 600(2006).
\bibitem{21} X. Jiang, G. Wu, J. Zhou, S. Wang, A. A. Tseng, Z. Du, Nanoscale Research Letters \textbf{6,} 518 (2011).
\bibitem{22} R. Garcia, R. V. Martinez, J. Martinez, Chem. Soc. Rev. \textbf{35,} 29 (2006).
\bibitem{23} R. Proksch, http://arxiv.org/pdf/1312.6933v1.pdf.
\bibitem{24} D. Andrienko, Cyclic Voltammetry, January 22,(2008).
\bibitem{25} A. N. Morozovska, E. A. Eliseev, S. V. Kalinin, Appl. Phys. Lett. \textbf{89,} 192901 (2006).
\bibitem{26} S. Jesse, B. Mirman, S. V. Kalinin, Appl. Phys. Lett. \textbf{89,} 022906 (2006).
\bibitem{27} S. Jesse, A. P. Baddorf, S. V. Kalinin, Appl. Phys. Lett. \textbf{88,} 062908 (2006).




\end{thebibliography}
\end{document}